\documentclass[twoside,reprint,superscriptaddress,amsmath,amssymb,aps]{revtex4-1}
\usepackage[latin9]{inputenc}
\usepackage{amsmath}
\usepackage{amssymb}
\usepackage{graphicx}

\usepackage{color}
\usepackage[english]{babel}
\usepackage[normalem]{ulem}


\begin{document}

\title{Self-diffusion and structural properties of confined fluids in dynamic coexistence}

\author{N. de Sousa}
\address{Departamento de F\'{i}sica
de la Materia Condensada and Condensed Matter Physics Center (IFIMAC), Universidad Aut\'{o}noma de Madrid, 28049 Madrid,
Spain.}

\author{J.J. S\'aenz}
\address{Donostia International Physics Center (DIPC), Paseo Manuel Lardizabal 4, 20018 Donostia-San Sebastian, Spain.}
\address{IKERBASQUE, Basque Foundation for Science, 48013 Bilbao, Spain.}

\author{Frank Scheffold}
\address{Department of Physics, University of Fribourg, Chemin du Mus\'{e}e 3,
CH-1700, Fribourg, Switzerland.}

\author{A. Garc\'{i}a-Mart\'{i}n}
\address{IMM-Instituto de Microelectr\'{o}nica de Madrid (CNM-CSIC), Isaac Newton
8, PTM, Tres Cantos, E-28760 Madrid, Spain}

\author{L. S. Froufe-P\'{e}rez}
\email{luis.froufe@unifr.ch, lfroufe@gmail.com}
\address{Department of Physics, University of Fribourg, Chemin du Mus\'{e}e 3,
CH-1700, Fribourg, Switzerland.}

\selectlanguage{english}%

\begin{abstract}
Self-diffusion and radial distribution functions are studied in a strongly confined Lennard-Jones fluid.
Surprisingly, in the solid-liquid phase transition region, where the system exhibits dynamic coexistence,
the self-diffusion constants are shown to present up to three-fold variations from solid to liquid phases at
fixed temperature, while the radial distribution function corresponding to both the liquid and the solid phases are essentially indistinguishable.
\end{abstract}

\pacs{82.60.Qr; 05.70.-a; 65.80.-g; 05.70.Fh}
\date{\today}
\maketitle

\section{Introduction}

The thermodynamics and molecular dynamics of gases, liquids, and solids
confined to small spaces can differ significantly from the bulk \cite{Pawlow:1909,Hill:1963}.
The confinement of a fluid in a region few times the particle diameter
induces density layering and solvation force oscillations \cite{Hansen:2006,Israelachvili:1991,bhushan1995nanotribology}
and can strongly modify the dynamical properties of the fluid \cite{Granick1991,Klein1995,Wales1996},
such as the diffusion of its constituents \cite{Erpenbeck:1991,Thompson1992,Gao1997,Mittal:2008,Matsubara:2012}.
The confinement also affects many other macroscopic properties of
the fluid \cite{Gelb:1999}, from capillary condensation \cite{evans1984theory,kober2010nanogeometry}
to melting/freezing phase transitions \cite{briant1975molecular,Buffat:1976,berry1984melting,Honeycutt:1987,berry1988solid,ercolessi1991melting,schmidt1998irregular,Baletto2005}.

For most liquids, the self-diffusion coefficient in highly confined
geometries can decrease (the viscosity can increase) by several orders
of magnitude with respect to the macroscopic bulk values \cite{Granick1991,Klein1995,Erpenbeck:1991,Thompson1992,Gao1997,Mittal:2008,Matsubara:2012}.
Although confinement strongly affects local structuring, the relationships
between self-diffusivity and thermodynamic quantities were found to
be, to an excellent approximation, independent of the confinement
\cite{Mittal:2008,Mittal:2006}, suggesting that thermodynamics can
be used to predict how confinement impacts dynamics \cite{Rosenfeld:1977}.
More recently, it has been shown that dynamic and equilibrium properties
have been explicitly related in supercooled and strongly confined
liquids \cite{Ingebrigtsen_PRL_2013}. 
In clusters, crystal nucleation (or the transition from liquid to solid) takes place spontaneously in supersaturated solutions. The size of the clusters is crucial to its evolution: if reaches a critical value, it grows; otherwise it will re-dissolve \cite{Auer:2001}.
According to classical nucleation theory, the transition is dominated by the surface free energy that accounts for the solid liquid interface.
However, in small clusters the surface free energy of the interface is so large that the system cannot afford coexistence between two different fases. As consequence, in equilibrium, the system jams from one phase to another and space coexistence in not possible.
Due to this impossibility of forming interfaces, the dynamics strongly differs from that observer in ether Van der Waals systems and/or hard spheres with colloidal systems \cite{Pham:2004}. 
These findings open an interesting
question about the nature of the self-diffusion near the freezing/melting
transition in confined geometries. In contrast with macroscopic systems,
for small clusters the transition does not take place at a well defined
temperature: there is a finite temperature range where solid and liquid
phases may coexist dynamically in time \cite{briant1975molecular,berry1984melting,Honeycutt:1987,berry1988solid,Labastie1990,Wales1994,Cleveland1994,schebarchov2005static},
i.e., observing the cluster over a long time, the cluster fluctuates
between being entirely solid or liquid.

Numerical simulations have been extensively used to analyze the size
dependence of the thermodynamic properties of confined fluids and
clusters \cite{briant1975molecular,Honeycutt:1987,berry1988solid,Quirke:1988,Polak:2006}.
Concerning the dynamics and size-dependence of self-diffusion in confined
fluids, most of the theoretical work have been focused on numerical
Molecular Dynamics (MD) simulations \cite{Beck1990,Erpenbeck:1991,Yeh:2004,Thompson1992,Gao1997,Mittal:2008,Matsubara:2012}.
Dynamic coexistence is not always observed in simulations \cite{Cleveland:1999}
but the observation of dynamic coexistence will of course depend on
the time scale on which dynamic coexistence occurs \cite{schebarchov2005static},
which can be very large depending on the magnitude of the energy barrier
separating the solid and liquid states of the cluster. Dynamic Monte
Carlo (DMC) simulations \cite{Binder1994} offer an alternative approach
that can be used to describe self-diffusion at large time scales \cite{Huitema:1999}
where both MD and DMC simulations reveals self-diffusion in confined
fluids as a thermal activated process \cite{Matsubara:2012}.

In this work we analyze and discuss the peculiar behavior of the self-diffusion
coefficients and radial distribution function, $g\left(r\right)$,
in a confined Lennard-Jones (L-J) fluid in the solid-liquid dynamic
coexistence region. We show that the spatial average of the self-diffusion
coefficients vary largely from liquid to the solid phase, providing
an unambiguous signature of the actual phase state. Interestingly,
we find that the $g\left(r\right)$ is essentially indistinguishable
between both phases. This indicates that the system is in an amorphous
solid phase rather than crystal-like. This finding is supported by
the observed split-second peak of $g(r)$ which is reminiscent of
the behavior observed in nearly jammed disordered hard-sphere (HS)
packings \cite{Donev_2005}. We shall term solid or
solid-like to such a phase throughout this paper.

It is worth emphasizing at this point that the interaction
potential is not hard-sphere although some similarities can be established
between HS and L-J systems. On the other hand, we consider spatially
averaged quantities. We shall show that, despite the possible spatial
dependence of the diffusion coefficients, spatially averaged quantities
already contain the signature of the the actual dynamic state of the
system.

\section{Lennard-Jones model}

More specifically, we study the self-diffusion coefficient of a medium
size (515 atoms) Lennard-Jones cluster confined in a spherical cavity
as a function of the temperature. In the liquid (fluid-like) phase,
just above the melting temperature, the self-diffusion coefficient
obtained from DMC numerical simulations follows the typical Arrhenius
behavior expected for thermal activated diffusion. In the coexistence
region, the self-diffusion randomly jumps between liquid-like to solid-like
reinforcing the relationship between dynamic and thermodynamic properties
even in this region. Although the confinement induces a strong anisotropy
of the pair-correlation functions of the fluid \cite{Nygard2012},
we find no significant differences in the average radial distribution
function between the two phases. Our results suggest that the direct
observation of dynamic coexistence could be accessible by experimental
approaches sensitive to self-diffusion by Nuclear Magnetic Resonance
\cite{kimmich1997nmr} or Fluorescence Correlation Spectroscopy \cite{Magne_Biopolimers_1974}
measurements for instance.

\subsection{Montecarlo simulations}

We start by studying a canonical ensemble of point particles interacting
through a L-J potential: 

\begin{equation}
V_{LJ}\left(r\right)=\varepsilon\left[\left(\frac{r_{m}}{r}\right)^{12}-2\left(\frac{r_{m}}{r}\right)^{6}\right],\label{eq:sSH_eq}
\end{equation}

where $\varepsilon$ is the depth of the potential well, $r$ is the
distance between particles and $r_{m}$ is the equilibrium distance.

The L-J fluid is confined inside a sphere. In order to consider a
high density in the system, the radius of the confining sphere is
chosen in such a way that a highly symmetric portion of a face centered
cubic (FCC) lattice fits the spherical volume. To have nearly relaxed
structures at zero temperature, the nearest neighbors distance of
the FCC lattice is chosen to be the equilibrium distance $r_{m}$.

Throughout this work, unless otherwise specified,
we consider $515$ confined point particles interacting through the
L-J potential given by eq. (\ref{eq:sSH_eq}), being the number density
$\rho\simeq1.07r_{m}^{-3}$. To have a clearer picture on how compact
this system is, we define an effective filling fraction $\phi$ by
considering that each particle effectively occupies a spherical volume
of radius $r_{m}/2$. The effective filling fraction of the system
under consideration is $\phi\simeq56\%$.

In order to generate a suitable statistical ensemble at fixed temperature,
we perform standard MC simulations using the canonical ensemble. We
depart from a crystalline structure and perform $10^{8}$ of MC steps
to thermalise the system. After this process an extensive MC sampling
is performed ($10^{5}$ configurations, each configuration obtained
after $10^{5}$ single-particle MC steps). Temperature and energy
is given in units of the potential well.

\subsection{Phase transitions in the system}

We determine the temperatures of the (isochore) phase transition in
the system by considering the specific heat (SH). The SH $C_{v}$
is obtained through the fluctuations of the internal energy\cite{Wang:2001jk}:

\begin{equation}
C_{v}\left(T\right)=\frac{\partial U\left(T\right)}{\partial T}=\frac{<E^{2}>-<E>^{2}}{T^{2}}.\label{eq:Cv}
\end{equation}

\begin{figure}[h]
\begin{centering}
\includegraphics[height=1\columnwidth,angle=-90]{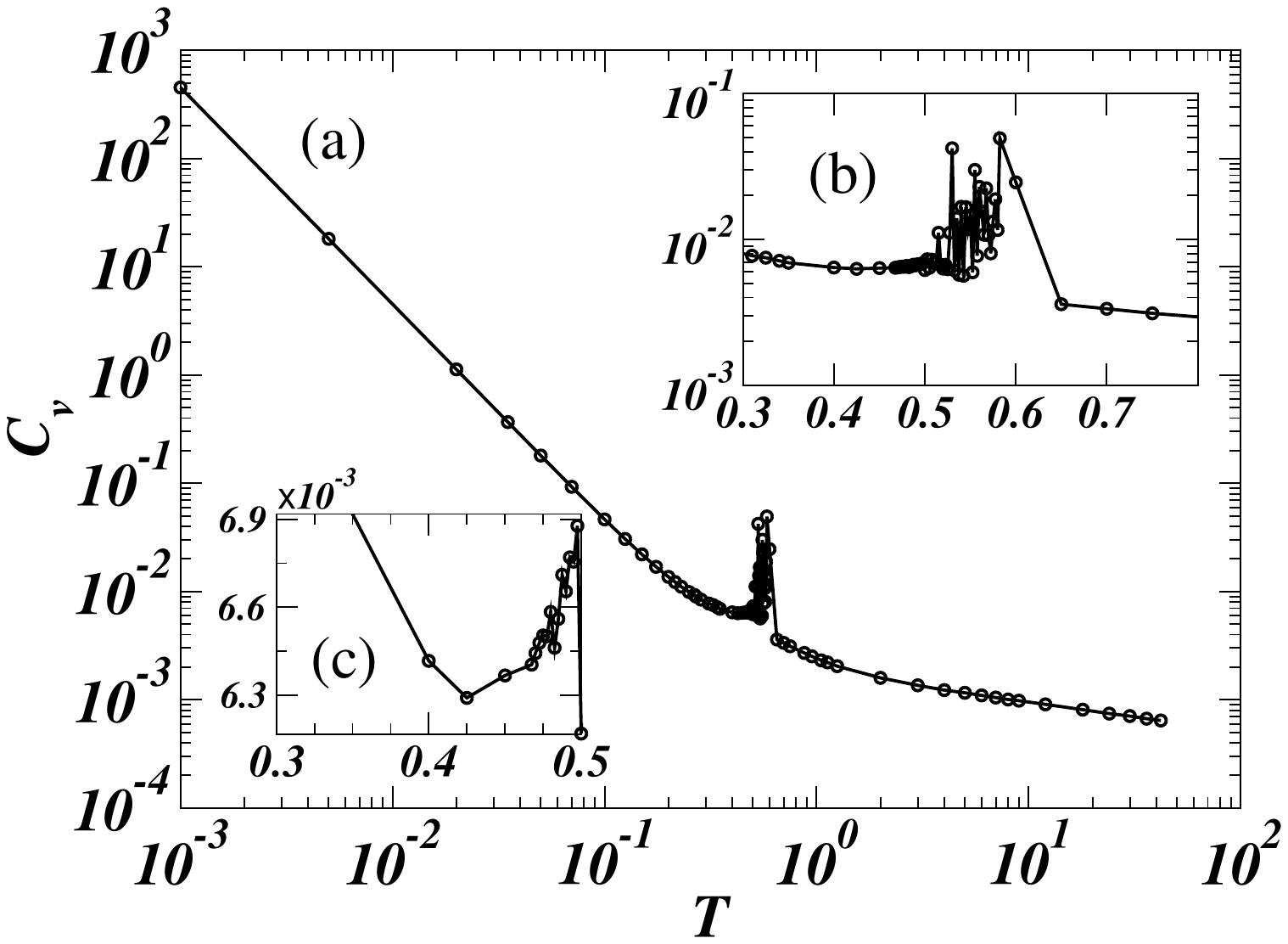} 
\par\end{centering}

\caption{\label{fig:specific_heat}Specific heat as function of temperature
for a confined L-J system with $N=515$ particles and an effective
filling fraction $\phi\simeq56\%$. Zooms of the specific heat is
represent in the box b) and c).}
\end{figure}

Considering the behavior of the specific heat as function of temperature,
as shown in fig. (\ref{fig:specific_heat}a), we observe a high and
narrow peak, which we ascribe to a first order phase transition for
$T\approx0.5$. Notice that in the phase transition region we have
relevant fluctuations, as can be observed in fig. (\ref{fig:specific_heat}c).
Also we observe a modification on SH for temperatures between $0.4\lesssim T\lesssim0.5$
(fig. \ref{fig:specific_heat}b), this feature in the SH might be
attributed to a pre-melting region.

In order to better describe the phase transitions
in the system, we also estimate the self-diffusion coefficient in
the system as a function of the temperature. To do so, the mean squared
displacement (MSD) $\left\langle \Delta R^{2}\right\rangle $ of particles
as function of the performed MC steps were fitted to a linear law.
In fig. (\ref{fig:massive_diffusion_coef}d) we show some representative
cases of this fitting procedure. From the slope the $\left\langle \Delta R^{2}\right\rangle $
vs. the number of Montecarlo steps the diffusion coefficient is extracted.
Since we take the averaged $\left\langle \Delta R^{2}\right\rangle $
considering all the particles at the same foot, we have a spatial
average of the diffusion coefficient. One might expect to find strong
inhomogeneities and anisotropy leading to an inhomogeneous diffusion
tensor instead of the averaged scalar values we obtain with our procedure.
Nevertheless, we shall see that this averaged diffusion constant suffices
to identify phase transitions and a dynamical phase switching regime
in our system.

\begin{figure}
\begin{centering}
\includegraphics[height=1\columnwidth,angle=-90]{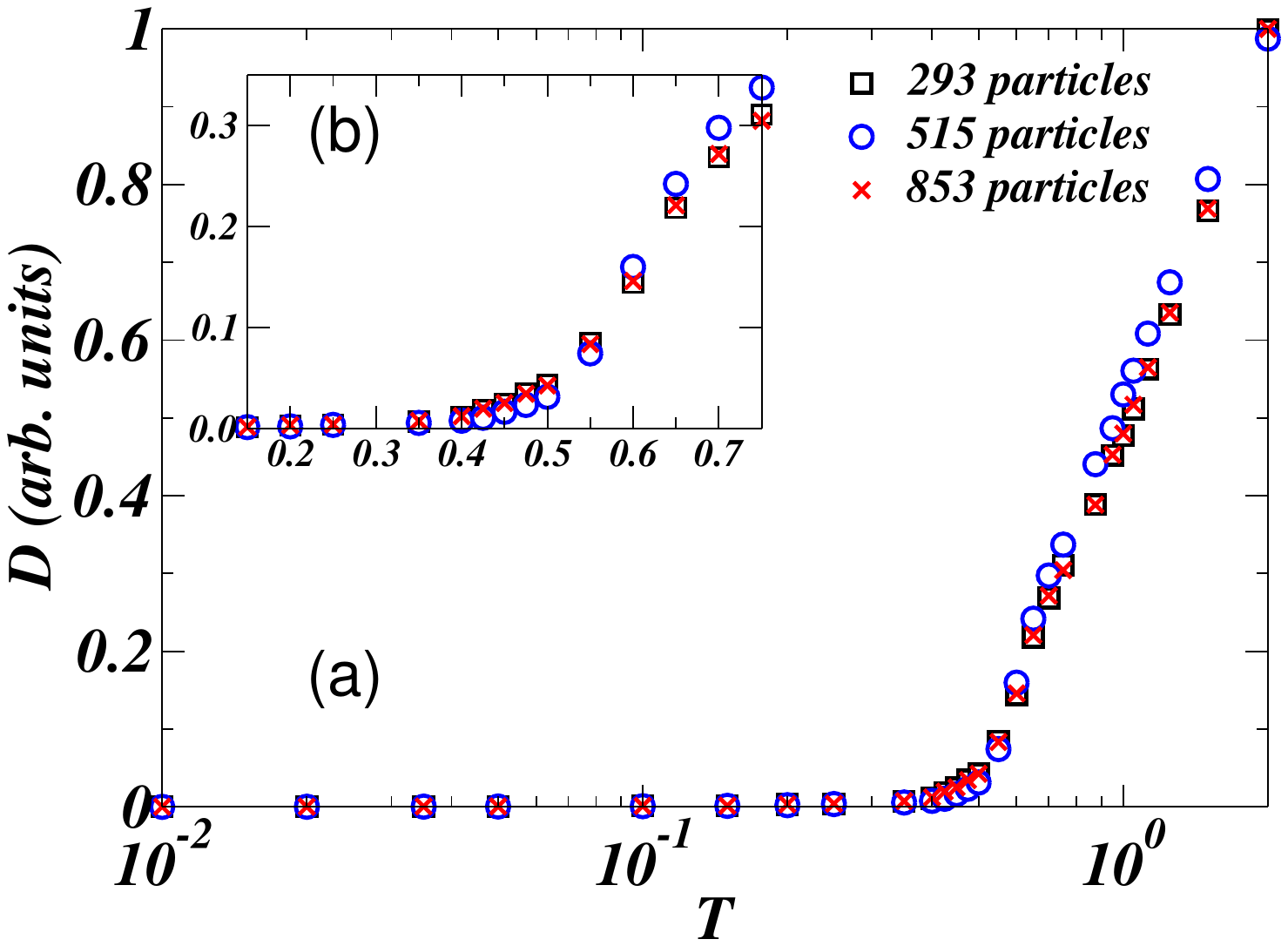} 
\par\end{centering}

\caption{(Color online) a) Diffusion coefficient as a function of the temperature,
for three different system sizes of the system at constant particle
density. b) Zoom of the same plot in the range $0.15<T<0.75$.\label{fig:diff_coef_1}}
\end{figure}

In fig. (\ref{fig:diff_coef_1}) we plot the diffusion coefficient
($D$) as a function of temperature for three different systems with
different number of particles and different volumes, while keeping
a constant number density. We observe that the diffusion coefficient,
for this scale, does not depend of the size of the system.

Three regions can be identified in fig. (\ref{fig:diff_coef_1}).
In the first region, for normalized temperatures $T\lesssim0.4$,
the structure is crystalline and diffusion is strongly inhibited.
This fact is compatible with a pure solid phase. The diffusion coefficient
grows with temperature at an approximately constant rate in the range
$0.4\lesssim T\lesssim0.5$ (See fig. \ref{fig:diff_coef_1}b). This
apparent increase in $D$ signals a pre-melting. It is worth noticing
that this region does not correspond to any remarkable feature in
the specific heat. The slope of the diffusion constant shows a strong
increase at about $T\simeq0.5$, this kink in the diffusion coefficient
curve corresponds to the peak in the specific heat.

In summary, we can establish a phase landscape in which, we identify
a pre-melting region that starts at $T\simeq0.4$, and a (solid-liquid)
phase transition at $T\simeq0.5$. In the following sections we shall
focus on the behavior of the system in the phase transition region.

\section{Phase switching}

\begin{figure}
\begin{centering}
\includegraphics[height=0.7\columnwidth]{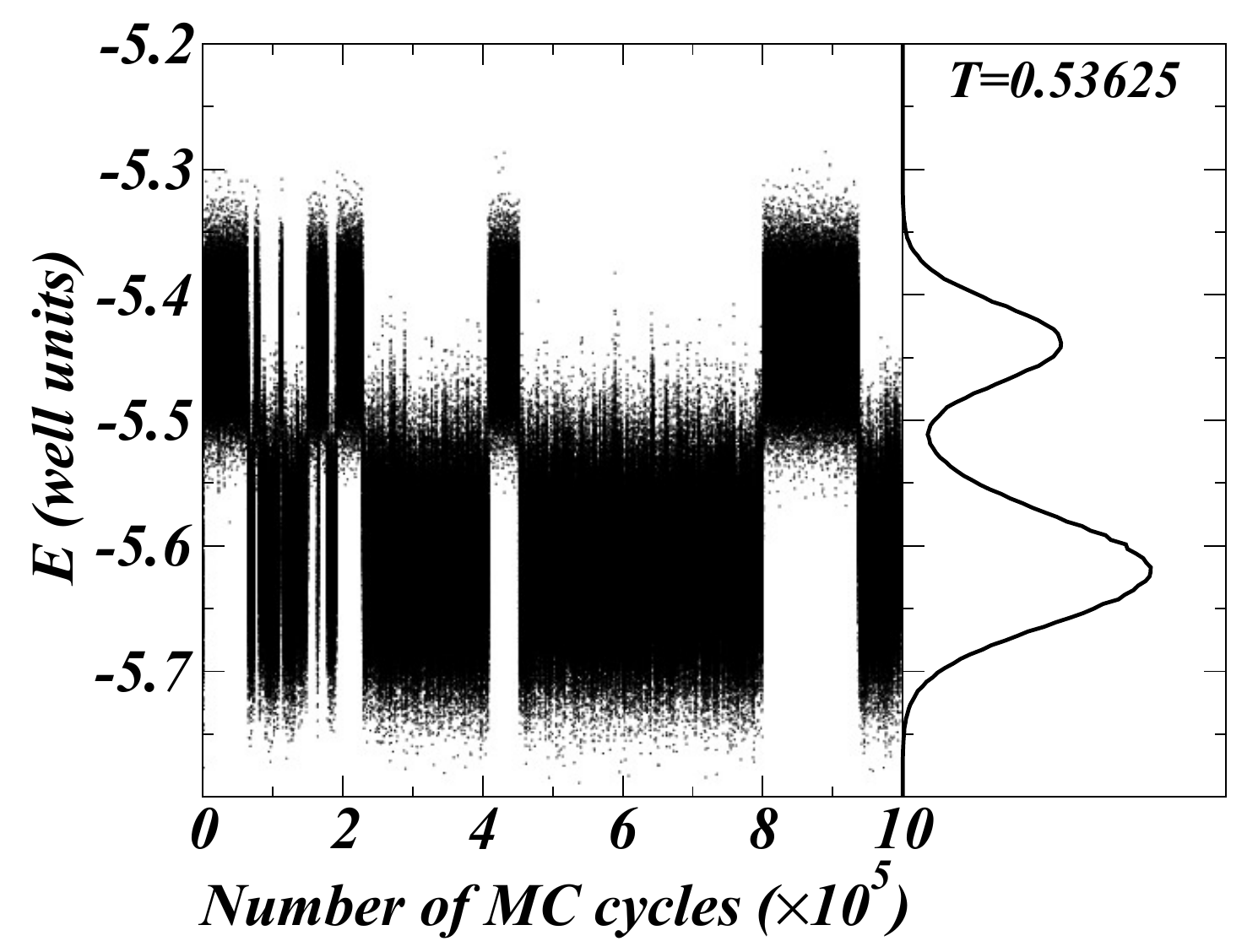} 
\par\end{centering}

\caption{\label{fig:energy_fluctuations} a) Energy sampling of a confined
L-J system during a full MC run at a temperature $T=0.53625$, corresponding
to a phase-switching region. b) Energy histogram obtained from the
MC sampling in a). }
\end{figure}

In fig. (\ref{fig:energy_fluctuations}) we represent the particle
energy as function of the MC steps for temperature $T=0.53625$, which
corresponds to a temperature in the phase transition region. The system
at this temperature oscillates between a lower and a higher value
of energy. This bistable energy behavior is the responsible for the
fluctuations in the SH. Despite the large number of MC steps used
in the sampling, we observe in fig. \ref{fig:energy_fluctuations}
that the number of high and low energy regions is relatively reduced.
Hence, if we calculate the SH through the energy fluctuations of internal
energy, large fluctuations due to finite sampling is expected as observed
in fig. (\ref{fig:specific_heat}c).

\begin{figure}[h]
\begin{centering}
\includegraphics[height=1.1\columnwidth]{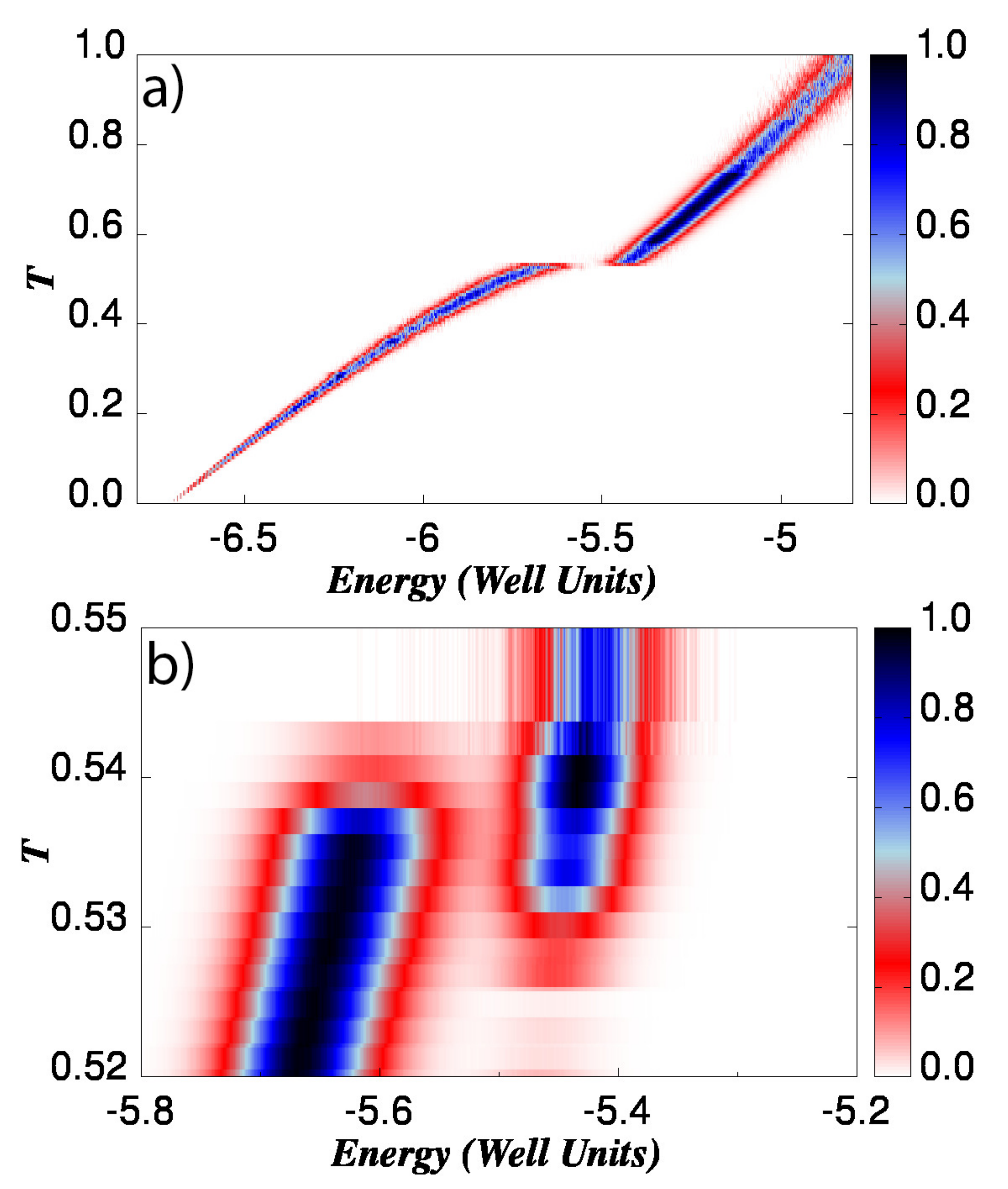} 
\par\end{centering}

\centering{}\caption{\label{fig:colormap_distrib}(Color online) a) Colormap showing the
energy distribution functions as a function of the temperature. b)
Zoom in the region corresponding to the solid-liquid phase transition.}
\end{figure}

Representing the internal energy histogram as function of the temperature,
shown in fig. (\ref{fig:colormap_distrib}), we can identify an energy
gap for temperatures at the phase transition. The transition between
solid and liquid is not smooth with phase coexistence between two
states. Instead, the system switches between this two phases, with
abrupt modifications in a small number of MC cycles. In the phase
transition, when the particles exhibit a low energy configuration,
the system is in the solid phase. For higher energies, the system
is in the liquid phase. 

One might expect to find intermediate energetic states
in the sampling. If both phases coexist the system might switch, or
smoothly evolve, among a multitude of different closely spaced energetic
states. Nevertheless in all the examined cases the system exclusively
switches between two well defined states. 

In fact one might expect phase coexistence for larger
systems, however, it seems that the size of the current system
is small enough to preclude phase coexistence. Apparently the system
is so small that a nucleation bubble fills the entire available volume. 

In order to better understand the geometrical and dynamical properties
of the system in the phase switching region, we observe that the system
remains in either the lower or the upper energy branches for a sufficient
amount of time (MC steps) to consider both the structural (pair correlation
function) or dynamical (self-diffusion constant) properties in well
defined phases.

\begin{figure}
\begin{centering}
\includegraphics[height=1\columnwidth]{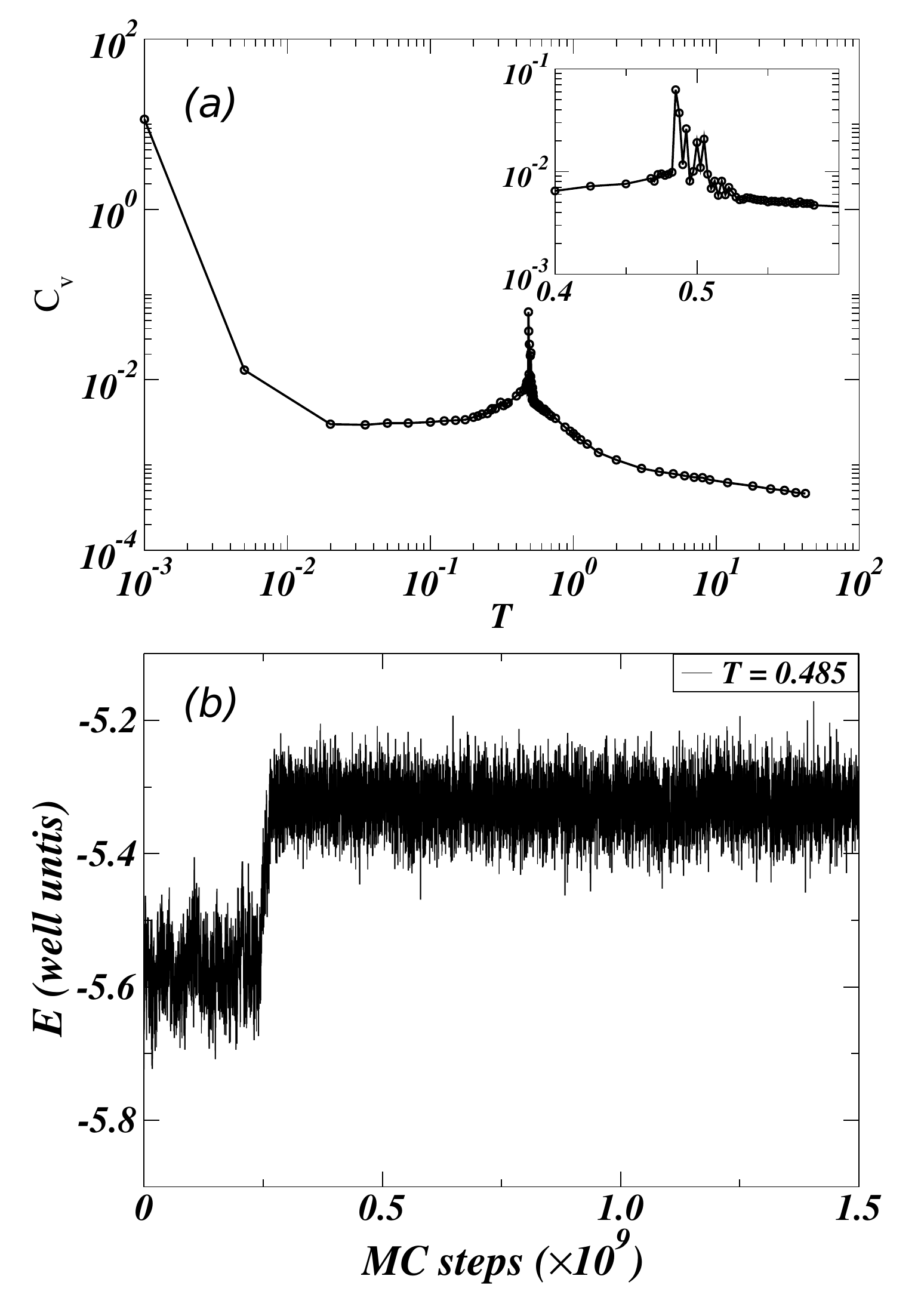} 
\par\end{centering}

\caption{\label{fig:filling_45_data}a) Specific heat as function of temperature
for a confined L-J system with $N=515$ particles and an effective
filling fraction $\phi\simeq45\%$. Zooms of the specific heat is
represent in the inset. b) Energy sampling for the same system during
a full MC run at a temperature $T=0.485$, corresponding to a phase-switching
region.}
\end{figure}

So far we focused our attention to a system with high
number density (effective filling fraction $\phi\simeq56\%$). Since
the interaction potential, although possessing a strong repulsive
core, is not a hard-core one, we expect that this behavior can be
maintained down to lower densities. We have performed several MC runs
on a system with an effective filling fraction $\phi=45\%$ obtaining
similar results for the $C_v$ as a function of temperature and a finite
phase transition region. In fig. (\ref{fig:filling_45_data}a), the
SH as a function of $T$ is represented. In analogy with previous results,
the SH curve shows different areas separated by a narrow peak at the
phase transition region . Again, the SH in this region strongly fluctuates
due to the finite sampling. 

In fig. (\ref{fig:filling_45_data}b) we plot an
energy sampling analogous to the one appearing in fig. (\ref{fig:energy_fluctuations}a).
In this case the temperature is $T=0.485$, corresponding to the maximum
of the SH peak. This energy sampling suggests that at this lower density,
the phase systems also might switch between two different dynamical
states. However the switching apparently happens at a much lower rate
than in the previous case. Notice that in this case we performed $\sim1.5\times10^{9}$
MC steps to detect one switching event in the energy, while in fig.
(\ref{fig:energy_fluctuations}a) several events were detected using
much less MC steps. Hence, although extensive simulations would be
required, we conjecture that the dynamical behavior presented in this
work might be found for any high enough density. The exact meaning
of high enough density can not be explicitly given with the available
data. In the remainder of this paper, we shall deal only with $\phi=0.56$
systems.

Using much larger systems might allow for the appearance
of more than two energy levels and diffusion constants. As a result,
the system would evolve with its size to a regime in which actual
phase coexistence instead of phase switching takes place.

\subsection{Dynamical properties in the switching regime}

\begin{figure}
\begin{centering}
\includegraphics[height=1\columnwidth]{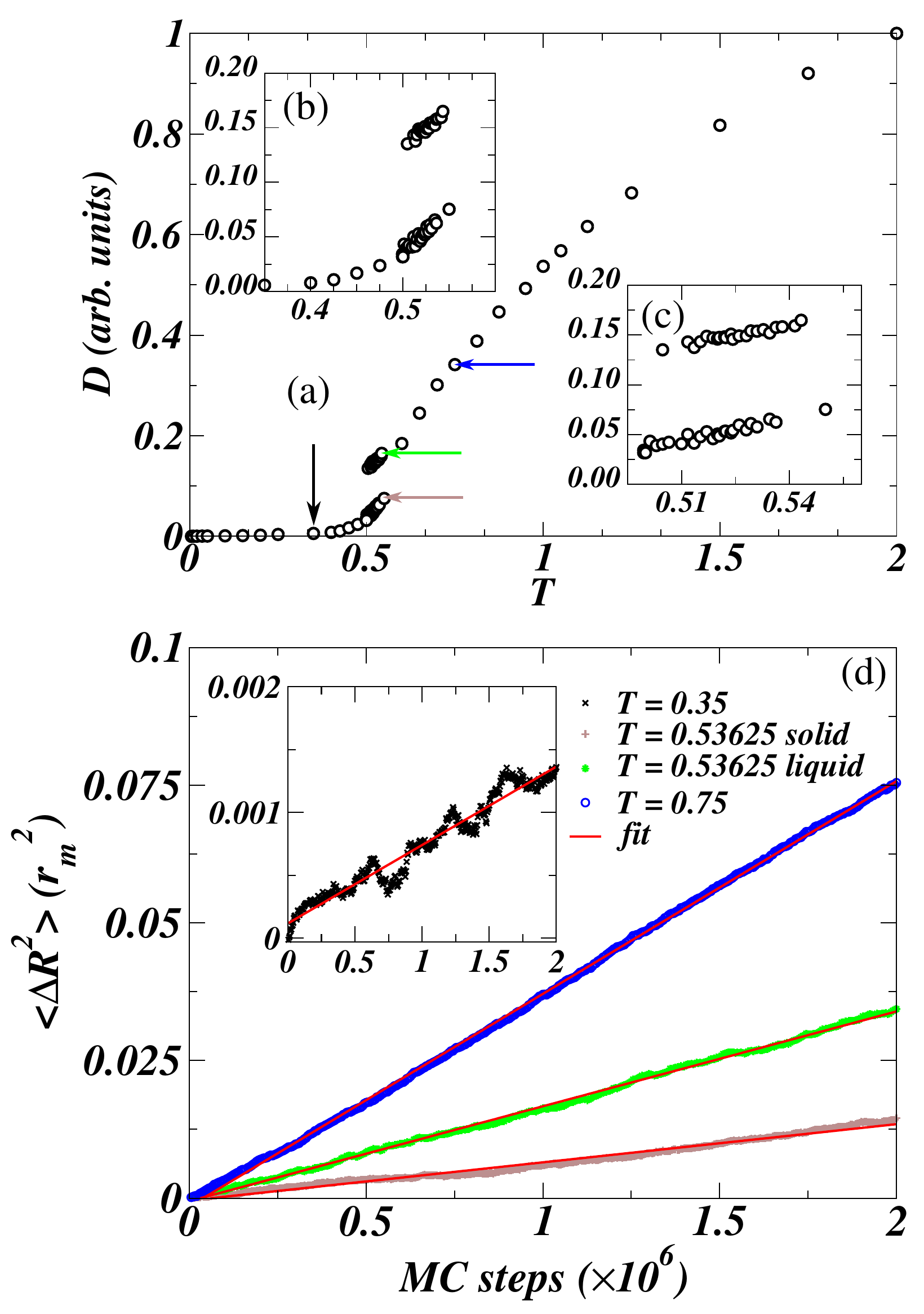} 
\par\end{centering}

\caption{\label{fig:massive_diffusion_coef} a) Self-diffusion
coefficient for a 515 particle system as a function of temperature.
Arrows indicate the points corresponding to data in panel d. b) Zoom
of the self-diffusion coefficient in the range $0.35\leq T\leq0.6$.
c) Self-diffusion coefficients as a function of temperature obtained
for the liquid phase (upper branch) and the solid phase (lower branch)
in the region of phase-switching. d) Mean squared displacement as
a function of MC steps for different data points shown in a) as indicated
by the arrows of the corresponding color. Red lines indicate the linear
fit from whose slope the diffusion coefficient is extracted.}
\end{figure}

Regarding dynamical properties, in fig. (\ref{fig:massive_diffusion_coef}a-c)
we plot the self-diffusion constant as a function of temperature much
in the same way as done in fig. \ref{fig:diff_coef_1}. In this case,
we have split the statistical ensemble in two different sets for temperatures
in the phase switching region, one corresponding to the high energy
branch (liquid phase), and the other one corresponding to lower energy
branch (solid phase). In fig. (\ref{fig:massive_diffusion_coef}c)
it appears evident that the diffusion coefficients corresponding to
both phases can differ by a large amount. In the case under study,
the diffusion constant differs by a factor 3 between phases at the
same temperature.

In fig. (\ref{fig:massive_diffusion_coef}d) we show
several examples of the evolution of the mean squared displacement
as a function of the number of MC steps. In all the examined cases,
the maximum root mean squared value is well below the radius of the
confining sphere, hence we do not expect saturation effects due to
confinement. Nevertheless, at higher temperatures and long simulation
times, the MSD tends to saturate after a purely diffusive region as
expected (not shown for the sake of brevity). 

\subsection{Geometrical properties in the switching regime}

\begin{figure}
\begin{centering}
\includegraphics[clip,height=0.9\columnwidth,angle=-90]{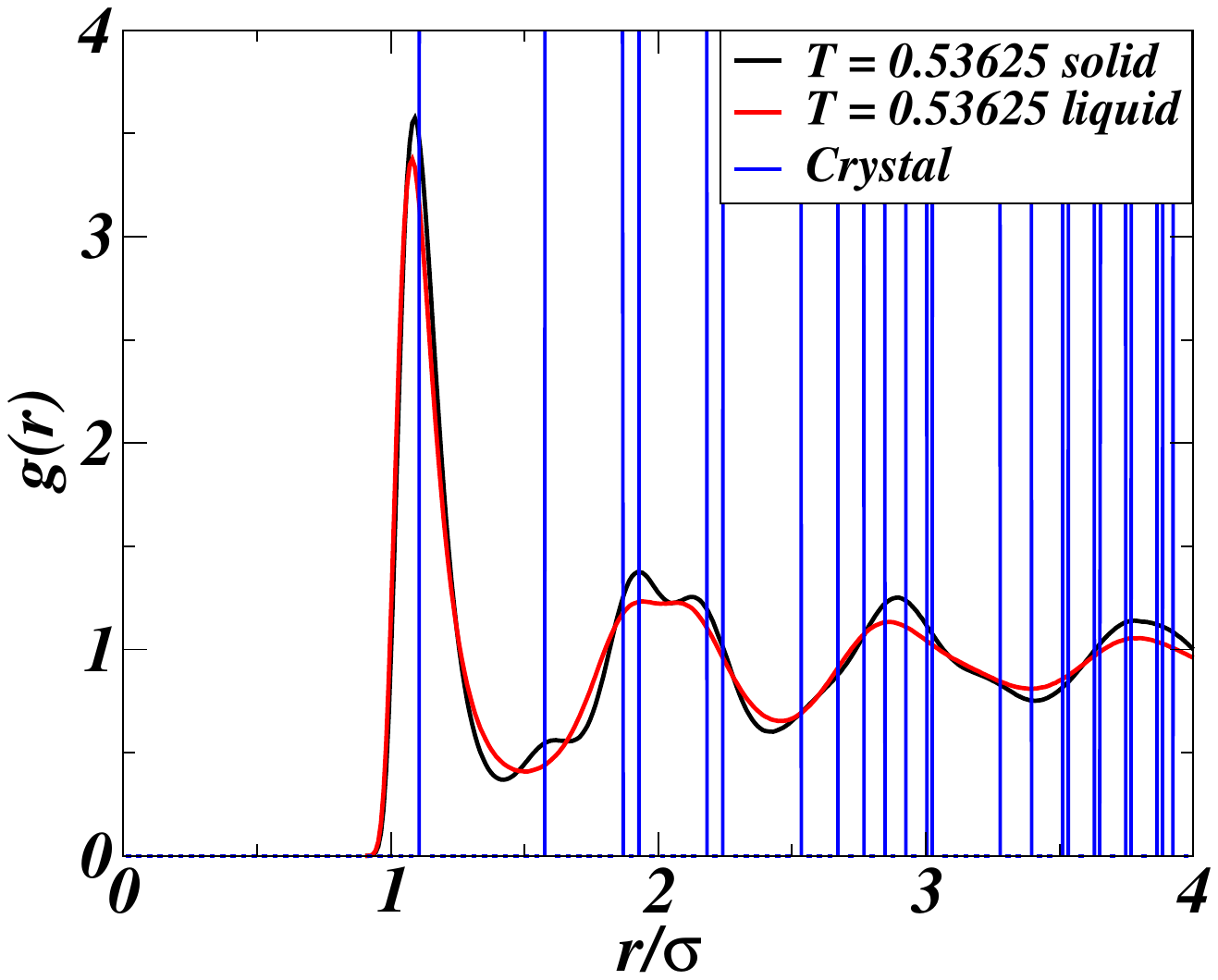} 
\par\end{centering}

\caption{\label{fig:Radial-distribution-function}(Color online) Radial distribution
function $g\left(r\right)$ obtained at a fixed temperature in the
phase switching region ($T\simeq0.53$) for both the solid (black
line) and liquid (red line) phases. Blue vertical lines correspond
to a perfect FCC crystal. }
\end{figure}

Regarding geometrical properties of both phases at the phase switching
region, we have studied the radial distribution function $g\left(r\right)$
\cite{Percus:1958wv,Wertheim:1963tb}. This function is defined as
the ratio of the average number density at a distance $r$ from one
particle to the averaged number density of an hypothetical, fully
uncorrelated, system. Hence, the radial distribution function describes
the correlation in the interparticle distance in the system. Again,
we can distinguish the statistical sampling in two sets associated
with upper and lower energy branches in the phase switching region.
Contrary to the behavior of the diffusion constants, the radial distribution
function in the upper and lower energy branches is very similar. In
fig. \ref{fig:Radial-distribution-function} we represent the $g\left(r\right)$
for the configurations at both the liquid and solid phase at a fixed
temperature. The only marked difference is the indicated split-second
peak of $g\left(r\right)$ which for bulk packings of hard spheres
is a known signature of a solid phase \cite{Donev_2005,Liu:2010}.
Other than that the radial distribution functions $g\left(r\right)$
remain nearly identical when the switching from solid to liquid phases
and a clear identification of different phases through structural
measurements is therefore much less sensitive than through dynamical
measurements (eg. self-diffusion constants) in strongly confined systems.

\section{Conclusion}

In summary, we have studied the self-diffusion in a strongly confined
Lennard-Jones system. For small clusters, of the order of a few hundreds
of particles, instead of phase coexistence the system present dynamic
phase switching between solid-like and liquid-like amorphous phases.
We found that the self-diffusion coefficient of the liquid-like phase
in the phase-switching region can be up to a factor of three larger
than the one associated to the solid phase. Interestingly, although
the radial distribution functions are nearly the same a split-second
peak is observed as a subtle structural signature of transient solid
phase.

\begin{acknowledgments}
The authors acknowledge Profs. Yuriy G. Pogorelov and Manuel Marqu\'{e}s
for useful and stimulating discussions. This research was supported by the Spanish Ministry of Economy and Competitiveness through grant MINIELPHO FIS2012-36113-C03. 
AGM acknowledges financial  support by the Spanish Ministry of Economy and Competitiveness (Contract Nos. MAT2011-29194-C02-01, MAT2014-58860-P)  and by the Comunidad de Madrid (Contract No. S2013/MIT-2740).
LSFP and FS acknowledge financial support by the Swiss National Science Foundation through the National Centre of Competence in Research Bio-Inspired Materials.
\end{acknowledgments}

\end{document}